\documentclass[aps,prd,twocolumn,nofootinbib,preprintnumbers,amssymb,amsfonts,amsmath,superscriptaddress,showpacs]{revtex4-1}
\usepackage{graphicx}
\usepackage{amsfonts}
\usepackage{amssymb}
\usepackage{amsmath}
\usepackage{hyperref}
\usepackage{slashed}
\usepackage{cancel}
\usepackage{enumerate}% http://ctan.org/pkg/enumerate
\usepackage{color}
\def\eq#1{{eq.~(\ref{#1})}}
\def\eqs#1#2{{eqs.~(\ref{#1})--(\ref{#2})}}

\def\Tr{\mbox{Tr}\,}

\def\hhref#1{\href{http://arxiv.org/abs/#1}{#1}} % in bibliography

\definecolor{oucrimsonred}{rgb}{0.6, 0.0, 0.0}
\definecolor{persianblue}{rgb}{0.11, 0.22, 0.73}
\definecolor{forestgreen}{rgb}{0.13,0.35,0.13}
 \hypersetup{colorlinks, citecolor=oucrimsonred, linkcolor=persianblue, urlcolor=oucrimsonred}
\newcommand{\be}{\begin{equation}}
\newcommand{\ee}{\end{equation}}
\newcommand{\bea}{\begin{eqnarray}}
\newcommand{\eea}{\end{eqnarray}}
\newcommand{\nn}{\nonumber}

\begin{document}
\preprint{CERN-TH-2016-007}
%%%%%%%%%%%%%%%%%%%%%%%%%%%%%%%%%%%%%%%%%%%%%%%%%%%%%%%%%%%  FRONT PAGE
\title[]{The breaking of the $SU(2)_L\times U(1)_Y$ symmetry\\
{\small The 750 GeV resonance at the LHC and perturbative unitarity}}
\date{\today}
\author{Marco Fabbrichesi}
\email{marco@sissa.it}
\affiliation{INFN, Sezione di Trieste, via Valerio 2, I-34137 Trieste, Italy}
\author{Alfredo Urbano}
\email{alfredo.leonardo.urbano@cern.ch}
\affiliation{Theoretical Physics Department, CERN, Geneva, Switzerland}
\begin{abstract}
\noindent  If the di-photon excess at 750 GeV hinted by the 2015 data at the LHC is explained in terms of a scalar resonance participating in the breaking of the electro-weak   symmetry, this resonance  must be accompanied by other scalar states for perturbative unitarity in vector boson scattering  to be preserved.  The simplest set-up consistent with perturbative unitarity and with the data of the di-photon excess is the  Georgi-Machacek model.

\end{abstract}

%\keywords{}
%\pacs{14.60.Pq, 14.80.Bn, 95.35.+d}
%%%%%%%%%%%%%%%%%%%%%%%%%%%%%%%%%%%%%%%%%%%%%%%%%%%%%%%%%%%%%%%%%%%
\maketitle
%%%%%%%%%%%%%%%%%%%%%%%%%%%%%%%%%%%%%%%%%%%%%%%%%%%%%%%%%%%%%%%%%%%

%%%%%%%%%%%%%%%%%%%%%%%%%%%%%%%%%%%%%%%%%%%%%%%
\section{Motivations}\label{eq:SectionI}
%%%%%%%%%%%%%%%%%%%%%%%%%%%%%%%%%%%%%%%%%%%%%%%%
Irrespective of whether it will stay or not---the recent excess in the  2015 LHC data with two  photons in the final state at invariant mass of about 750 GeV~\cite{750}  reminds us that even after the discovery of the Higgs boson we may still  not know all the details of the breaking of the electro-weak (EW) symmetry. 

Let us interpret  the LHC di-photon excess as a new scalar resonance. 

The simplest (although perhaps least interesting) possibility is that this  resonance takes no part in the breaking of the EW symmetry. In this case, it is possible to reproduce the di-photon excess by coupling the resonance---in a generic fashion---to extra scalar or fermionic degrees of freedom (see, for instance, \cite{Knapen:2015dap} and \cite{Franceschini:2015kwy}). If this is the case, the rationale of such new physics is  bound to remain rather mysterious and we might be  justified in thinking  that it  would be for the best if the  di-photon excess were to disappear from the new data in 2016.  

On the other hand, if this
 resonance takes part in the EW symmetry breaking, its existence would tell us something new about such a mechanism, in particular that it is not realised  by the vacuum expectation value (VEV) of the Higgs boson alone. Moreover---and more importantly for the present work---the presence of such a state  necessarily affects the high-energy behavior of the theory: to the extent that the  perturbative unitarity of vector boson scattering is to be preserved, such a resonance cannot come by itself or with arbitrary couplings~\cite{Grinstein:2013fia}. 

Let us classify states  after symmetry breaking according to their properties under custodial $SU(2)_C$ and take the new  resonance  to be a singlet.  There are  two possibilities. This custodial singlet either 
\begin{itemize}
\item comes from one or more doublets  (this choice leads to the two Higgs doublet model (2HDM)~\cite{Branco:2011iw}
 and related constructions) and its coupling to the gauge bosons is fixed by gauge invariance to combine with that of the Higgs boson to cancel the unitarity violating growth with the center-of-mass (CM) energy; or 
\item its coupling to the gauge bosons  does not combine with that  of the Higgs boson as to cancel the unitarity violations,  and we must   also  include a quintuplet of custodial $SU(2)_C$---the only scalar with a contribution in the high-energy amplitudes of the opposite sign  with respect to  that of the Higgs boson and other singlets~\cite{Lee:1977eg}---in order for unitarity to be preserved. 
\end{itemize}

The inclusion of a custodial singlet resonance arbitrary coupled to the gauge bosons therefore leads  naturally to the Georgi-Machacek (GM) model ~\cite{Georgi:1985nv}---the simplest model to contain a  custodial quintuplet and in which symmetry breaking is achieved by three scalar fields: one doublet (with hypercharge 1/2) and two triplets (with hypercharges 1 and 0).

If neither of the above  is the case, perturbative unitarity cannot be preserved and the singlet resonance must belong to a non-perturbative regime. This would imply the exciting discovery of  a new interaction that is strong at the EW scale. A fit of the di-photon excess in terms of a non-perturbative resonance is possible and has been already discussed in the literature (for instance, see \cite{Franceschini:2015kwy}). 

In this paper we expand on the reasoning above. We discuss to what extent a singlet resonance can take part in the EW symmetry breaking and still belong to a perturbative regime in which reliably computations can be performed. The  GM model seems to emerge as the simplest  model satisfying these requirements  that also explains the di-photon excess at the LHC for a realistic choice of its parameters.

%%%%%%%%%%%%%%%%%%%%%%%%%%%%%%%%%%%%%%%%%%%%%%%%
%\vskip 1.3em
\subsection{Perturbative unitarity}
%%%%%%%%%%%%%%%%%%%%%%%%%%%%%%%%%%%%%%%%%%%%%%%%
 Perturbative unitarity limits the possible models in which the leading  orders of perturbation theory are expected to be a reliable guide to physics~\cite{Lee:1977eg}. If perturbative unitarity is satisfied, EW interactions are described by a renormalisable gauge theory and the strength of the interactions among the particle remains weak at all energies. If this is not the case, unitarity is recovered by the inclusion of higher order terms; these, however, cannot be small and a non-perturbative regime is entered.
 
The requirement of perturbative unitarity is stated in terms of partial-wave amplitudes $a_J(s)$ where the amplitude of vector boson scattering is
 \be
 a_{VV}(s,t) = 16 \pi \sum (2J+1) a_J(s) P_J(\cos \theta) \, ,
 \ee
 and  $s$ and $t$ are the Mandelstam variables.
 Unitarity requires that 
 \be
| a_0(s)| <1 \, .
\ee

In general, the partial-wave amplitude in vector boson scattering is given by
 \be
 a_J(s) =  A \left(\frac{\sqrt{s}}{m_W} \right) ^4 +  B \left(\frac{\sqrt{s}}{m_W} \right) ^2 +  C \left(\frac{\sqrt{s}}{m_W} \right) ^0 \, , \label{eq:unitarity}
 \ee
with terms growing as the fourth power and the square of the CM energy,  and a constant, respectively. 
 $A$ vanishes by gauge invariance that implies $g_{4V}=g^2_{3V}$. $B$ vanishes in the standard model (SM) because of the Higgs boson $h$ contribution and the relationship
 \be
 m_V^2 g_{4V} - \frac{3}{4} m_V^2 g^2_{3V} = \frac{1}{4} g^2_{hVV}
 \ee 
among the couplings (with self-explanatory notation).  The constant terms in $C$ sets a limit on the Higgs boson mass in the SM and on the masses of other states in its extensions. 
 
 If there are more singlets, for instance two: $H_1$ and $H_1'$,   their couplings must satisfy
 \be
  m_V^2 g_{4V} - \frac{3}{4} m_V^2 g^2_{3V} = \frac{1}{4} \left( g^2_{H_1VV} + g^2_{H'_1VV} \right) \label{eq:2hdm}
 \ee
in order for the coefficient $B$ in \eq{eq:unitarity} to vanish.  This is realised in the 2HDM and variations of the same.
 
 The other possibility is to have a negative contribution: this can only come from a   quintuplet (see \cite{Lee:1977eg} and  \cite{Alboteanu:2008my}) of custodial $SU(2)_C$. In fact, for  interactions
 \be
 \frac{g_{H_1'} v}{2} H_1 \, \Tr D_\mu \Sigma^\dag D^\mu \Sigma  \label{eq:1}
 \ee
 and
 \be
  - \frac{g_{H_5} v}{2} H_5 \, \left[ D_\mu \Sigma^\dag  D^\mu \Sigma
 - \frac{\sigma^{aa}}{6} \Tr D_\mu \Sigma^\dag D^\mu \Sigma \right] \label{eq:5}
 \ee
 between the longitudinal components of the vector boson fields $\Sigma = \exp \left[ -i/v \sum \sigma^a \pi^a \right]$ and the singlet in \eq{eq:1} and quintuplet in \eq{eq:5},
 the amplitudes for singlet scalars are always
 \be
 a(s,t)|_{H_1'} = -\frac{g_{H_1}^2}{v^2} \frac{s^2}{s-m_H^2} \label{eq:tree1}
 \ee
 with the same sign as the Higgs boson, while
 \bea
 a(s,t)|_{H_5} & =&   -\frac{g_{H_5}^2}{v^2} \left[  \frac{t^2}{t-m_{H_5}^2} \right.  \nn \\
 & & \left. + \frac{u^2}{u-m_{H_5}^2} - \frac{2}{3} \frac{s^2}{s-m_{H_5}^2} \right] 
 \label{eq:tree5}\, ,
 \eea
gives a (repulsive) negative contribution. 

Considering the limit $s \gg m_W^2, m_{H_1}, m_{H_1}' m_{H_5}$---and having the Higgs boson contribution already cancel the contribution  from the vector bosons to the coefficient $B$ in \eq{eq:unitarity}---an exact cancellation between eq.~(\ref{eq:tree1}) and eq.~(\ref{eq:tree5}) requires
\begin{equation}
\frac{5}{6}g_{H_5}^2 = g_{H_1'}^2\, .
\end{equation}
As shown below, such a cancellation, and the unitarity of the theory, are automatically implemented in the GM model.

 %%%%%%%%%%%%%%%%%%%%%%%%%%%%%%%%%%%%%%%%%%%%%%%%
%\vskip 1.3em
\section{The first possibility: the 2HDM}\label{sec:2HDM}
%%%%%%%%%%%%%%%%%%%%%%%%%%%%%%%%%%%%%%%%%%%%%%%% 
 
 The first possibility considered in the introduction section is the simplest: perturbative unitarity is maintained by having the scalar resonance coupling at a special value fixed  by gauge invariance (see \eq{eq:2hdm}). 
 
 This would be the first choice in trying to incorporate the resonance within a model. Unfortunately, the parameters of the 2HDM model must be pushed to  rather unrealistic values in order to accomodate the di-photon data~\cite{Altmannshofer:2015xfo}. These values are particularly worrisome in the light of the required size of the  the Yukawa couplings, the renormalized values of which  bring the theory into a non-perturbative regime~\cite{Son:2015vfl}. 
 
 We therefore consider the other case discussed in section I.

 %%%%%%%%%%%%%%%%%%%%%%%%%%%%%%%%%%%%%%%%%%%%%%%%
%\vskip 1.3em
\section{The GM model}
%%%%%%%%%%%%%%%%%%%%%%%%%%%%%%%%%%%%%%%%%%%%%%%%
The GM model contains a complex $SU(2)_L$ doublet field $\phi$ ($Y = 1$), a real 
triplet field $\xi$ ($Y = 0$), and a complex  $SU(2)_L$ triplet field $\chi$ ($Y = 2$).
The scalar content of the theory can be organised 
in terms of the $SU(2)_L \otimes SU(2)_R$ symmetry, and we define the following multiplets
\bea\label{eq:GMScalars}
\Phi_{ (\textbf{2},\textbf{2})} &\equiv&
\left(
\begin{array}{cc}
\phi^{0*}  & \phi^+     \\
 \phi^- & \phi^0      
\end{array}
\right), \\
\quad \Delta_{ (\textbf{3},\textbf{3})}&\equiv& 
\left(
\begin{array}{ccc}
 \chi^{0*} & \xi^+  & \chi^{++}   \\
 \chi^- & \xi^0  & \chi^+  \\
 \chi^{--} & \xi^{-}  & \chi^0   
\end{array}
\right),
\eea
whose VEVs are 
\be
\langle \Phi \rangle = \frac{v_\phi}{\sqrt{2}}\, \hat I_{2\times 2} \quad \mbox{and} \quad \langle \Delta \rangle = \frac{v_\Delta}{\sqrt{2}} \, \hat I_{3\times 3} \, ,
\ee
 with
$v_\phi^2 + 8 v_\Delta^2 = v^2 = 1/\sqrt{2}\,  G_F \simeq (246 \, \mbox{GeV})^2$.   The VEVs of the two triplets must be the same in order to preserve custodial $SU(2)_C$.

The doublet and the two triplet states can be written in components:
\bea
\phi &=&
\left(
\begin{array}{c}
   \phi^+  \\
 (v_{\phi} + \phi^0_{r} + \imath \phi^0_{i})/\sqrt{2}   
\end{array}
\right)~,
 \\
\xi &=&
\left(
\begin{array}{c}
 \xi^+   \\
 v_{\Delta} + \xi^0    \\
  \xi^-
\end{array}
\right)~, \\
\chi &=&
\left(
\begin{array}{c}
  \chi^{++}   \\
  \chi^+   \\
  v_{\Delta} + ( \chi^0_{r} + \imath \chi^0_{i})/\sqrt{2}  
\end{array}
\right)~,
\eea
with $\phi^- = -(\phi^+)^*$, $\xi^- = -(\xi^+)^*$, $\chi^- = -(\chi^+)^*$.

The most general potential that conserves  $SU(2)_C$ is given by
\bea\label{eq:Potential}
V(\Phi, \Delta) & =& \frac{\mu_2^2} \Tr \Phi^\dag \Phi + \frac{\mu_3^2}{2} \Tr \Delta^\dag \Delta  \nn \\
& & + \lambda_1 \left[  \Tr \Phi^\dag \Phi \right]^2 + \lambda_2  \Tr \Phi^\dag \Phi \, \Tr \Delta^\dag \Delta \nn \\
& & + \lambda_3 \Tr \Delta^\dag \Delta \Delta^\dag \Delta +   \lambda_4 \left[\Tr \Delta^\dag \Delta \right]^2  \nn\\
& & - \lambda_5 \Tr \left( \Phi^\dag \sigma^a \Phi \sigma^b  \right) \, \Tr  \left(\Delta^\dag t^2 \Delta t^b \right) \nn \\
& &  - M_1  \Tr \left( \Phi^\dag \tau^a \Phi \tau^b \right) (U \Delta U^\dag)_{ab}\nn  \\
& & - M_2 \Tr \left( \Delta^\dag t^a \Delta t^b \right) (U \Delta U^\dag)_{ab} \, ,
\eea 
where $\tau$ and $t$ are the $SU(2)$ generators in the doublet and triplet representation respectively, and $U$ a matrix that rotates $\Delta$ into the Cartesian basis.

From the (canonically normalised) kinetic terms 
\begin{equation}
\mathcal{L}_{\rm kin} = |D^{(\phi)}_{\mu}\phi|^2 + \frac{1}{2}|D^{(\xi)}_{\mu}\xi|^2 + |D^{(\chi)}_{\mu}\chi|^2 \, ,
\end{equation} 
we can read the interactions with the EW gauge bosons. 
Considering the neutral components of the scalar fields in eq.~(\ref{eq:GMScalars}),
a direct computation gives
\begin{eqnarray}\label{eq:Basic}
\mathcal{L}_{\rm kin} &\supset&
(v_{\phi} + \phi^0_{r})^2\left(
\frac{g^2}{4}W_{\mu}^+W^{-,\mu} + \frac{g^2 + g^{\prime\,2}}{8}Z_{\mu}Z^{\mu}
\right)\nonumber \\
&+& (v_{\Delta} + \xi^0)^2
\left(
g^2 W_{\mu}^+W^{-,\mu} 
\right) \\
&+& 
(\sqrt{2}\, v_{\Delta} + \chi^0_{r})^2\left(
\frac{g^2}{2}W_{\mu}^+W^{-,\mu} + \frac{g^2 + g^{\prime\,2}}{2}Z_{\mu}Z^{\mu} \nn
\right)\,.
\end{eqnarray}
The imaginary part of $\phi$ and $\chi$ does not interact with the EW gauge bosons as a consequence of $CP$ invariance.
The gauge boson masses are  given by
\begin{equation}
m_W^2 \equiv \frac{g^2}{4}(v_{\phi}^2 + 8 v_{\Delta}^2 )~,~~~~
m_Z^2 \equiv \frac{g^2 + g^{\prime\,2}}{4}(v_{\phi}^2 + 8v_{\Delta}^2)~.
\end{equation}

Under $SU(2)_C$ we have the group representations
$(\textbf{2},\textbf{2}) \sim \textbf{1} \oplus \textbf{3}$, and $(\textbf{3},\textbf{3}) \sim \textbf{1} \oplus \textbf{3} \oplus \textbf{5}$.
One of the two triplets is unphysical, since it represents the Goldstone bosons eaten by the EW gauge bosons. 
Accordingly, the GM model has ten physical degrees of freedom: 
two $SU(2)_C$ singlets $H_1^0$, $H_1^{0^{\prime}}$ (the Higgs and the additional scalar resonance), one $SU(2)_C$ triplet $(H_3^+, H_3^0, H_3^-)$ and 
one $SU(2)_C$ quintuplet $(H_5^{++}, H_5^+, H_5^0, H_5^-, H_5^{--})$.

If compared with the setup envisaged in section~\ref{eq:SectionI}, 
the spectrum of the GM model has one additional scalar triplet.
However, the triplet $H_3$ does not interact with the EW gauge bosons.

The mass eigenstates in terms of gauge eigenstates are
\begin{eqnarray}
H_5^{++} &=& \chi^{++}~,\nn \\
H_5^{+} &=& (\chi^+ - \xi^+)/\sqrt{2}~,\nonumber\\
H_5^{0} &=&  (2\xi^0 - \sqrt{2}\chi^0_{r} )/\sqrt{6}~,\nonumber\\
H_3^+ &=& \cos\theta_H(\chi^+ + \xi^+)/\sqrt{2} - \sin\theta_H\phi^+,\nonumber\\
H_3^0 &=& \imath (-\cos\theta_H \chi^0_{i} + \sin\theta_H\phi^0_{i})\, ,\nonumber\\
H_1^0 &=& \phi^0_{r}~,\nonumber\\
H_1^{0^{\prime}} &=& (\sqrt{2} \chi^0_{r} + \xi^0)/\sqrt{3}~.
\end{eqnarray}

From the Lagrangian in eq.~(\ref{eq:Basic}) we find the physical couplings
\begin{eqnarray}\label{eq:Canc1}
\mathcal{L}_{\rm kin} &\supset& \cos\theta_H\frac{H_1^0}{v}\left(
2m_W^2W_{\mu}^+W^{-,\mu} + m_Z^2 Z_{\mu}Z^{\mu}
\right)\\
&+& \frac{2\sqrt{2}}{\sqrt{3}}\sin\theta_H 
\frac{H_1^{0^{\prime}}}{v}
\left(
2m_W^2W_{\mu}^+W^{-,\mu} + m_Z^2 Z_{\mu}Z^{\mu}
\right)\nonumber \\
&+& \frac{2}{\sqrt{3}}\sin\theta_H \frac{H_5^{0}}{v}\left(
m_W^2W_{\mu}^+W^{-,\mu} - m_Z^2 Z_{\mu}Z^{\mu}
\right)~,\nonumber
\end{eqnarray}
where
the doublet-triplet mixing angle is given by
\begin{equation}\label{eq:VEVs}
\tan\theta_H \equiv 2 \sqrt{2}\,  \frac{v_{\Delta}}{v_{\phi}}~.
\end{equation}

As far as the charged interactions are concerned, we find, in the $g^{\prime} \to 0$ limit,
\begin{equation}\label{eq:Canc2}
\mathcal{L}_{\rm kin} \supset -2\sin\theta_H\frac{m_W m_Z}{v}H_5^+W_{\mu}^-Z^{\mu} + h.c.~.
\end{equation}
From the interactions in \eqs{eq:Canc1}{eq:Canc2} 
we have
\begin{equation}
g_{H_1^0 VV}^2   \equiv \cos^2\theta_H \quad \mbox{and} \quad g_{H_1^{0 '}}^2 \equiv \frac{8}{3}\sin^2\theta_H
\ee
for the singlets, and
\be
g_{H_5}^2 \equiv 2\sin^2\theta_H\,,
\end{equation}
for the quintuplet. The cancellation of the coefficient $B$ in the vector boson scattering amplitude
follows from
\begin{equation}
1- g_{H_1^0 VV}^2 - g_{H_1^{0' }}^2 + \frac{5}{6}g_{H_5}^2 = 0\,.
\end{equation}

\subsection{Mass spectra and couplings}

After EW symmetry breaking, a mixing between the neutral singlet scalar states $H_1^0$ and $H_1^{0^{\prime}}$ is generated.
The corresponding mass matrix is 
\begin{equation}\label{eq:MixingMatrix}
\mathcal{M}^2 =
\left(
\begin{array}{cc}
\mathcal{M}_{11}^2  & \mathcal{M}_{12}^2    \\
\mathcal{M}_{12}^2  &    \mathcal{M}_{22}^2  
\end{array}
\right)~,
\end{equation}
with 
\begin{eqnarray}
\mathcal{M}_{11}^2 &=& 8\lambda_1 v_{\phi}^2~,\\
\mathcal{M}_{12}^2 &=&
\frac{\sqrt{3}}{2}v_{\phi}\left[
-M_1 + 4\left(
2\lambda_2 -\lambda_5
\right)v_{\Delta}
\right]~,\nonumber \\
\mathcal{M}_{22}^2 &=& \frac{M_1v_{\phi}^2}{4v_{\Delta}} -6M_2v_{\Delta}
+8(\lambda_3 +3\lambda_4)v_{\Delta}^2~.\nonumber
\end{eqnarray}
The mass matrix can be easily diagonalized by introducing the physical states
\begin{equation}\label{eq:Higgs}
h = c_{\alpha}H_1^0 - s_{\alpha}H_1^{0^{\prime}}~,~
H = s_{\alpha}H_1^0 + c_{\alpha}H_1^{0^{\prime}}~,
\end{equation}
where $\alpha$ is a mixing angle and we used the short-hand notation $c_{\alpha}\equiv \cos\alpha$,
$s_{\alpha}\equiv \sin\alpha$ from which 
$\alpha = \pm \sin^{-1}[(1-c_{2\alpha})/2]$. The mass eigenvalues are 
\begin{equation}\label{eq:MassHiggs}
 2m_{h,H}^2 =
\mathcal{M}_{11}^2 +\mathcal{M}_{22}^2 \mp
\sqrt{
\Delta^2
}~,
\end{equation}
with $\Delta^2 \equiv (\mathcal{M}_{11}^2 - \mathcal{M}_{22}^2)^2 + 4(\mathcal{M}_{12}^2)^2$.
The mixing angle is defined by 
\be
s_{2\alpha} = \frac{2\mathcal{M}_{12}^2}{(m_H^2 - m_h^2)}\, . 
\ee

The masses of the custodial triplet and quintuplet are given by
\begin{eqnarray}\label{eq:HeavyMasses}
m_{H_3}^2 &=&  \left(
\frac{M_1}{4v_{\Delta}} + \frac{\lambda_5}{2}
\right)v^2~,\\
m_{H_5}^2 &=& \frac{M_1}{4v_{\Delta}}v_{\phi}^2 +12M_2v_{\Delta} + \frac{3}{2}
\lambda_5v_{\phi}^2 + 8\lambda_3v_{\Delta}^2~.
\end{eqnarray}
Neglecting loop-induced mass splitting, the mass is degenerate within the same custodial multiplet.

As a consequence of the rotation in eq.~(\ref{eq:Higgs}) and the ratio of VEVs in eq.~(\ref{eq:VEVs})
the Higgs couplings with gauge bosons and fermions are modified with respect to the corresponding SM values.
One finds
\begin{eqnarray}
g_{hW^+W^-} &=& -\frac{g^2}{6}\left(
8\sqrt{3}s_{\alpha}v_{\Delta} - 3c_{\alpha}v_{\phi}
\right)~,\\
g_{hf\bar{f}} &=& -\frac{\imath m_f}{v}\frac{c_{\alpha}}{\cos\theta_H}~,
\end{eqnarray}
with $g_{hW^+W^-}= c_W^2g_{hZZ}$.

 %%%%%%%%%%%%%%%%%%%%%%%%%%%%%%%%%%%%%%%%%%%%%%%%
%\vskip 1.3em
\section{Fitting the 750 GeV di-photon excess}
%%%%%%%%%%%%%%%%%%%%%%%%%%%%%%%%%%%%%%%%%%%%%%%%

There exists a number of constraints that the parameters of the GM model must satisfy in order to reproduce the observed di-photon excess while, at the same time, not be in violation of other known observables.

First of all, for the model to be consistent, its parameters must 
\begin{itemize}
\item[$\circ$] {\it Satisfy perturbative unitarity}.
Perturbative unitarity on the $2\to 2$ scalar field scattering amplitudes provides 
a set of stringent constraints on the parameters of the scalar potential~\cite{Hartling:2014zca}:
\begin{eqnarray}
\sqrt{P_{\lambda}^2 + 36\lambda_2^2} + |6\lambda_1 +7\lambda_3 + 11\lambda_4| &<& 4\pi~,\\
\sqrt{Q_{\lambda}^2 +\lambda_5^2} + |2\lambda_1 -\lambda_3 +2\lambda_4 | &<& 4\pi~,\\
|2\lambda_2 + \lambda_4|   &<& \pi~,\\
|\lambda_2 - \lambda_5|   &<& 2\pi~,
\end{eqnarray}
with $P_{\lambda} \equiv 6\lambda_1 -7\lambda_3 -11\lambda_4$,
$Q_{\lambda} \equiv 2\lambda_1 +\lambda_3 -2\lambda_4$. 
In addition, we also have
\begin{equation}\label{eq:25}
\lambda_2 \in \left(-\frac{2}{3}\pi, \frac{2}{3}\pi\right)~,~~~~
\lambda_5 \in \left(-\frac{8}{3}\pi, \frac{8}{3}\pi\right)~.
\end{equation}

\item[$\circ$] {\it Have a potential bounded from below}.
This requirement restricts $\lambda_{3,4}$ in the following interval 
\begin{equation}\label{eq:34}
\lambda_3 \in \left(-\frac{1}{2}\pi, \frac{3}{5}\pi\right)~,~~~~
\lambda_4 \in \left(-\frac{1}{5}\pi, \frac{1}{2}\pi\right)~.
\end{equation}
\end{itemize}

In addition, we must verify that, for each choice of parameters, known experimental constraints are satisfied.  These are:

\begin{itemize}

\item[$\circ$] {\it Modification of the SM Higgs couplings}.
Higgs coupling measurements~\cite{ATLAS_CMS}
strongly constrained the allowed values of $v_{\Delta}$ and $\alpha$.

\item[$\circ$] {\it Electroweak precision tests}. 
The presence of additional scalar states, charged under the EW symmetry, 
generates a non-zero contribution to the $S$ parameter~\cite{Hartling:2014xma}. 

\end{itemize}

\begin{figure*}[!htb]
\minipage{0.5\textwidth}
  \includegraphics[width=.75\linewidth]{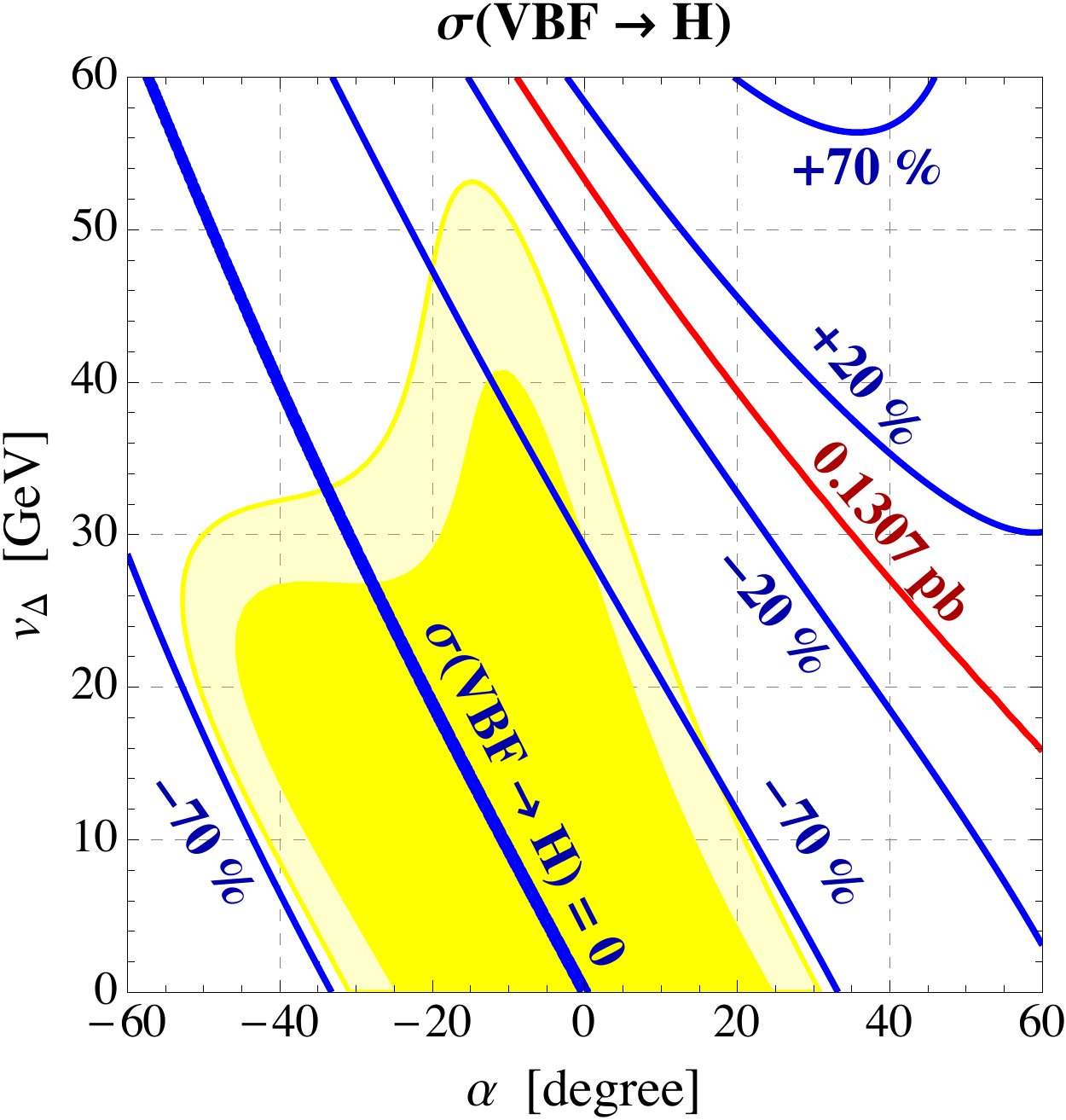}
\endminipage\hfill
\minipage{0.5\textwidth}
  \includegraphics[width=.75\linewidth]{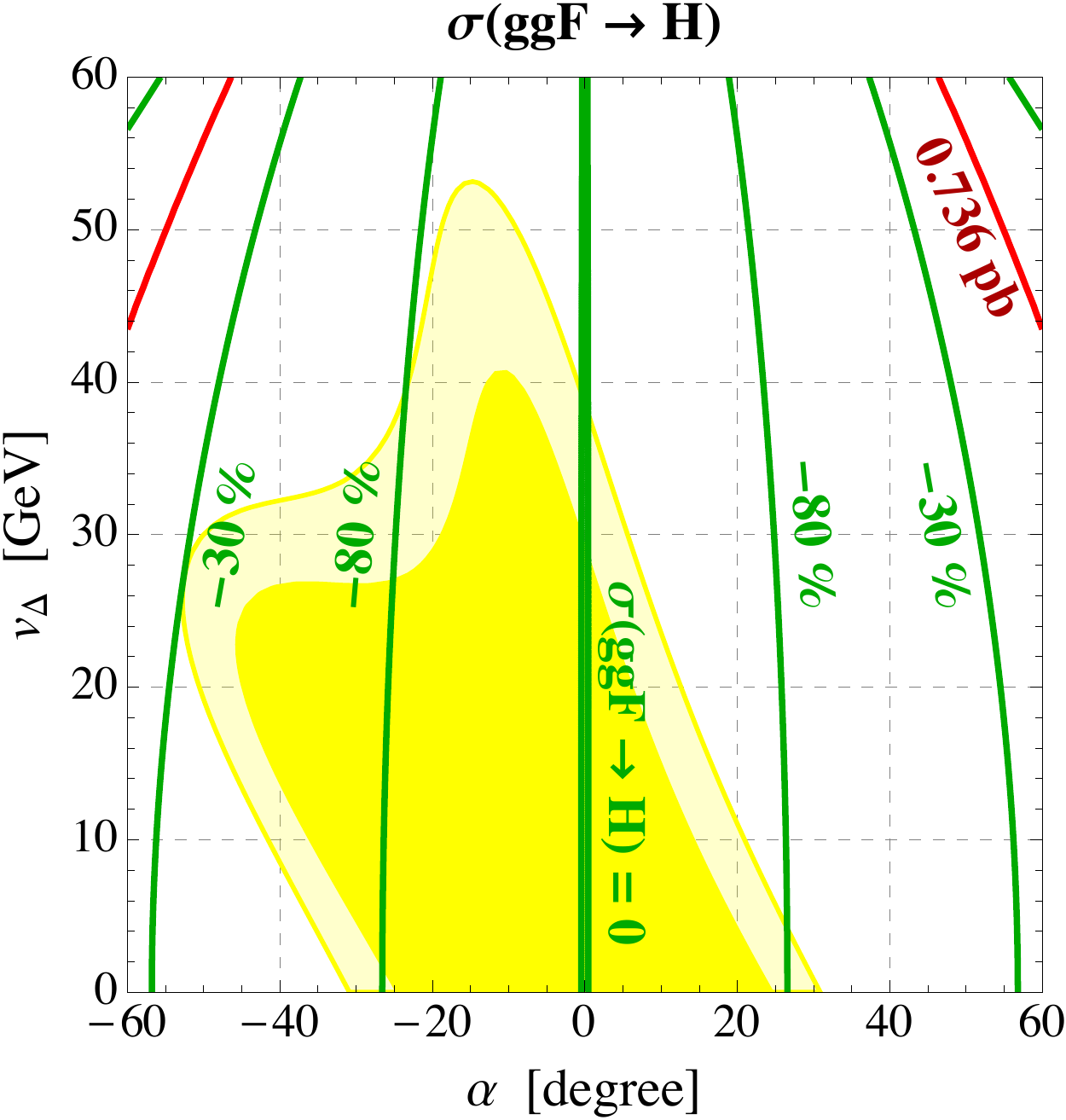}
\endminipage
 \caption{ \textit{
 Contours of production cross section for the scalar resonance $H$ via VBF (left panel) and ggF (right panel) at $\sqrt{s} = 13$ TeV
 in the two-dimensional plane $(\alpha, v_{\Delta})$.
In both cases the red line marks the production cross sections for a SM Higgs boson with $m_h = 750$ GeV,
that are $\sigma({\rm VBF} \to h)_{m_h = 750\,{\rm GeV}} \simeq 0.1307$ pb
and $\sigma({\rm ggF} \to h)_{m_h = 750\,{\rm GeV}} \simeq 0.736$ pb.
The yellow regions represent the $1$- and $2$-$\sigma$ confidence level (darker and lighter yellow, respectively)
allowed by the Higgs coupling measurements at the LHC.
  }}
 \label{fig:Production}
\end{figure*}

In order to  explore the model, we perform a parameter scan by  proceeding as follows:
\begin{itemize}
\item[\it 1.] The lightest state $h$ is the physical Higgs boson, with $m_h = 125.09$ GeV, while we identify 
the second mass eigenstate $H$ with the new resonance at $m_H = 750$ GeV.
Eq.~(\ref{eq:MassHiggs}) can be inverted, and one can fix two parameters of the scalar potential.
We solve eq.~(\ref{eq:MassHiggs}) for $\lambda_1$ and $M_1$;
\item[\it 2.] The parameter $\lambda_{2,3,4,5}$ are randomly generated within the intervals in 
eqs.~(\eqs{eq:25}{eq:34}); for each quadruplet, we check that the unitarity constraints are satisfied;
\item[\it 3.] The remaining parameters $v_{\Delta}$ and $M_2$ are randomly generated within the intervals $v_{\Delta} \in (0,50)$ GeV,
$|M_2| \in (1,10^4)$ GeV. The VEV $v_{\phi}$ is given by $v_{\phi} = \sqrt{v^2 - 8v_{\Delta}^2}$; 
\item[\it 4.] For each sample of values the mass matrix in eq.~(\ref{eq:MixingMatrix})---and hence the mixing angle $\alpha$---and the mass eigenstates in eq.~(\ref{eq:HeavyMasses}) can be computed;
\item[\it 5.] As a final step in our Monte-Carlo generation, we check that the values of $v_{\Delta}$ and $\alpha$ are consistent 
with the Higgs coupling measurements at the $2$-$\sigma$ level. 
Following~\cite{Falkowski:2013dza}, we perform a two-parameter 
$\chi^2$ fit of the most recent ATLAS and CMS measurements~\cite{ATLAS_CMS}.
We show in fig.~\ref{fig:Production} the corresponding $1$- and $2$-$\sigma$ confidence level contours in the plane $(\alpha, v_{\Delta})$. 

We also check that the 
correction to the $S$ parameter is within 3-$\sigma$ of the  LEP-I and LEP-II fit of the EW precision observables. In fig.~\ref{fig:Mixing} we show the constraint from the EW parameter $S$ on  the scan of the parameters $v_\Delta$ and $\alpha$ of the GM model.

\end{itemize}
 %%%%%%%%%%%%%%%%%%%%%%%%%%%%%%%%%%%%%
%%%%%%%%%%%%%%%%%%%%%%%%%%%%%%%%%%%%%

Having set the scope and range of the parameter scan, we are now in the position to discuss the fit of the di-photon excess.

\subsection{Production cross section}

The mixing with the Higgs boson in eq.~(\ref{eq:Higgs}) 
and the presence of a non-zero VEV $v_{\Delta}$
automatically allows for $H$ production via both Vector Boson Fusion (VBF) and gluon fusion (ggF).
The former is triggered by tree-level $H$ couplings with the EW gauge bosons, the latter at one loop 
by $H$ coupling to SM fermions, with the top quark providing the most sizable contribution.

The relevant couplings are 
\begin{eqnarray}
g_{HW^+W^-} &=& \frac{g^2}{6}\left(
8\sqrt{3}c_{\alpha}v_{\Delta} + 3s_{\alpha}v_{\phi}
\right)~,\\
g_{Ht\bar{t}} &=& -\frac{\imath m_t}{v}\frac{s_{\alpha}}{\cos\theta_H}~.
\end{eqnarray}

The $H$ production cross section can be straightforwardly obtained by rescaling 
the production cross section of a SM Higgs with $m_h = 750$ GeV.
At $\sqrt{s} = 13$ TeV we have $\sigma({\rm VBF} \to h)_{m_h = 750\,{\rm GeV}} \simeq 0.1307$ pb and 
$\sigma({\rm ggF} \to h)_{m_h = 750\,{\rm GeV}} \simeq 0.736$ pb~\cite{Xsec}, and the rescaling is simply given by 
\begin{eqnarray}
\sigma({\rm VBF} \to H) &=& (c_V^H)^2\times \sigma({\rm VBF} \to h)_{m_h = 750\,{\rm GeV}}~,\nonumber \\
\sigma({\rm ggF} \to H) &=& (c_F^H)^2\times \sigma({\rm ggF} \to h)_{m_h = 750\,{\rm GeV}}~,
\end{eqnarray}
where 
\begin{eqnarray}
c_V^H &=& \frac{1}{3}\left[
\frac{8\sqrt{3}c_{\alpha}v_{\Delta} + 3s_{\alpha}v_{\phi}}{v}
\right]~,\\
c_F^H &=& \frac{v s_{\alpha}}{\sqrt{v^2 - 8v_{\Delta}^2}}~.
\end{eqnarray}

The rescaled cross sections
crucially depend on the values of $v_{\Delta}$ and $\alpha$.
In fig.~\ref{fig:Production} we show contours 
of constant VBF (left panel, blue lines) and ggF (right panel, green lines) $H$ production compared with the reference values of the SM Higgs with $m_h = 750$ GeV (red lines).
As clear from the plot, in the allowed region of the $(\alpha, v_{\Delta})$ plane
we always observe a reduction if compared with the SM case.

In addition to VBF and ggF, we also include---following~\cite{Fichet:2015vvy}---production via photon fusion ($\gamma\gamma$F)
for inelastic,  partially  elastic  and elastic  collisions.

\subsection{Total decay width and di-photon decay}

The di-photon signal strength at $\sqrt{s} = 13$ TeV is given by
\begin{eqnarray}\label{eq:SignalStrength}
\mu_H &=& [
\sigma({\rm ggF}\to H) + \sigma({\rm VBF} \to H)
]\times {\mathcal{BR}}(H \to \gamma\gamma) \nonumber \\
&+& 10.8\,{\rm pb}\,\left(\frac{\Gamma_H}{45\,{\rm GeV}}\right)\times [{\mathcal{BR}}(H \to \gamma\gamma)]^2~,
\end{eqnarray}
where the last line accounts for production via $\gamma\gamma$F~\cite{Fichet:2015vvy}.

 Given the preliminary status of the experimental analysis, we do not perform any complicated fit.
On the contrary, the purpose of this section is to check whether the GM model can account for a di-photon signal strength 
of the order of few fb, that is the order of magnitude suggested by present data. 
As discussed in section~\ref{sec:2HDM}---a positive answer is anything but trivial in weakly coupled theories (in particular without invoking the presence of extra vector-like fermions with either large multiplicities, electric charge or Yukawa couplings)
and would be a remarkable result if achieved in the GM model.

In order to evaluate eq.~(\ref{eq:SignalStrength}) we need to compute 
the total decay width of the singlet, $\Gamma_H$, and the di-photon decay width.

At the tree level, 
$H$ predominantly decays---as far as the SM final states are concerned---into $W^+W^-$, $ZZ$, $t\bar{t}$ and $hh$. 
The corresponding decay widths can be computed rescaling those of the SM Higgs boson.
We find
\begin{eqnarray}\label{eq:Decay}
\Gamma_{VV}^{(H)} &=& \frac{G_{\mu}m_H^3 (c_V^H)^2\delta_V}{16\sqrt{2}\pi}
\sqrt{1-4x_V}(1-4x_V + 12x_V^2)~,\nonumber\\
\Gamma_{f\bar{f}}^{(H)} &=& \frac{G_{\mu}N_C m_H m_f^2 (c_F^H)^2}{4\sqrt{2}\pi}\left(
1-\frac{4m_f^2}{m_H^2}
\right)^{3/2}~,\nonumber\\
\Gamma_{hh}^{(H)} &=& \frac{g_{hhH}^2}{32\pi m_H}\sqrt{1-\frac{4m_h^2}{m_H^2}}~,
\end{eqnarray}
where $\delta_{V=W,Z} = 2(1)$, $x_V = m_V^2/m_H^2$, $G_{\mu} = 1/(\sqrt{2}v)^{1/2}$. The trilinear scalar coupling is~\cite{Hartling:2014zca}
\begin{eqnarray}
g_{hhH} &=& 24\lambda_1c_{\alpha}^2s_{\alpha}v_{\phi}
+ 8\sqrt{3}c_{\alpha}s_{\alpha}^2v_{\Delta}(\lambda_3 +3\lambda_4)\nn
 \\
&+& 2\left[
\sqrt{3}c_{\alpha}v_{\Delta}(3c_{\alpha}^2-2) + s_{\alpha}v_{\phi}(1-3c_{\alpha}^2)
\right](2\lambda_2 -\lambda_5)\nonumber\\
&-& \frac{\sqrt{3}}{2}M_1c_{\alpha}(3c_{\alpha}^2 - 2) - 4\sqrt{3}M_2c_{\alpha}s_{\alpha}^2~.
\end{eqnarray}
The singlet 
$H$ can also decay into the custodial 
triplet and quintuplet if the corresponding channels are kinematically allowed. 
If $m_H > m_{H_5}/2$ ($m_H > m_{H_3}/2$), the new decay channels are $\Gamma_{H_5^+H_5^-}^{(H)}$, 
$\Gamma_{H_5^{++}H_5^{--}}^{(H)}$, $\Gamma_{H_5^0H_5^0}^{(H)}$ ($\Gamma_{H_3^+H_3^-}^{(H)}$, $\Gamma_{H_3^0H_3^0}^{(H)}$).
The decay widths can be computed as in eq.~(\ref{eq:Decay}), and the relevant couplings are~\cite{Hartling:2014zca}:
\bea \label{eq:5}
g_{HH_5^0H_5^0} & =&  8\sqrt{3}(\lambda_3 
 + \lambda_4)c_{\alpha}v_{\Delta}  \nn \\
& & + (4\lambda_2 + \lambda_5)s_{\alpha}v_{\phi} + 2\sqrt{3}M_2c_{\alpha}~,
\eea
with $g_{HH_5^0H_5^0} = g_{HH_5^+H_5^-} = g_{HH_5^{++}H_5^{--}}$, 
and
\begin{eqnarray}\label{eq:3}
g_{HH_3^0H_3^0} &=& 64\lambda_1 s_{\alpha}\frac{v_{\Delta}^2v_{\phi}}{v^2}
+\frac{8v_{\phi}^2 v_{\Delta}}{\sqrt{3}v^2}c_{\alpha}(\lambda_3 + 3\lambda_4) \nn \\
&-& \frac{2\sqrt{3}M_2 v_{\phi}^2}{v^2}c_{\alpha} +  \frac{16v_{\Delta}^3c_{\alpha}}{\sqrt{3}v^2}
(6\lambda_2 + \lambda_5) \nonumber \\
&+& \frac{4v_{\Delta}M_1}{\sqrt{3}v^2}(c_{\alpha}v_{\Delta} + \sqrt{3}s_{\alpha}v_{\phi})
+ \frac{s_{\alpha}v_{\phi}^3}{v^2}(4\lambda_2 - \lambda_5) \nonumber \\
&+&
\frac{8\lambda_5v_{\Delta}v_{\phi}}{\sqrt{3}v^2}(c_{\alpha}v_{\phi} +\sqrt{3}s_{\alpha}v_{\Delta})~,
\end{eqnarray}
with $g_{HH_3^0H_3^0} = g_{HH_3^+H_3^-}$.

Finally, H can decay into a vector boson plus a custodial triplet scalar.
If $m_H > m_W + m_{H_3}$ and $m_H > m_Z + m_{H_3}$ the 
corresponding decay channels are $\Gamma^{(H)}_{W^{\pm}H_3^{\mp}}$ and $\Gamma^{(H)}_{Z H_3^{0}}$.
We find
\begin{equation}
\Gamma^{(H)}_{VH_3} = \frac{|g_{HVH_3}|^2 m_V^2}{16\pi m_H}
\lambda\left(
\frac{m_H^2}{m_V^2}, \frac{m_{H_3}^2}{m_V^2}
\right)
\lambda^{1/2}\left(
\frac{m_V^2}{m_H^2}, \frac{m_{H_3}^2}{m_H^2}
\right)~,
\end{equation}
where the kinematic function $\lambda$ is
$\lambda(x,y) = (1-x-y)^2 - 4xy$. 
The relevant couplings are
\begin{eqnarray}
g_{HZH_3^0} &=& \frac{i\sqrt{2} g}{\sqrt{3}c_W}
\left( \frac{c_{\alpha} v_{\phi}}{v} - \frac{\sqrt{3} v_{\Delta}s_{\alpha}}{v}
\right)~,\\
g_{HW^{\pm}H_3^{\mp}} &=&
-\frac{\sqrt{2} g}{\sqrt{3}}\left(
\frac{\sqrt{3}s_{\alpha}v_{\Delta}}{v}  - \frac{c_{alpha}v_{\phi}}{v}
\right)~.
\end{eqnarray}

The sum of the tree-level decay widths reconstruct the total width $\Gamma_H$.

%We are now in the position to reconstruct the di-photon decay width.
The loop-induced di-photon decay width 
for the scalar singlet $\mathcal{H} = h,H$ is therefore
\begin{eqnarray}\label{eq:diphoton}
\Gamma_{\gamma\gamma}^{(\mathcal{H})} &=& \frac{G_{\mu}\alpha^2 m_{\mathcal{H}}^3}{128\sqrt{2}\pi^3}\left|
\sum_f N_CQ_f^2 g_{\mathcal{H}f\bar{f}}A_{1/2}^{\mathcal{H}}(\tau_f) \right.\nonumber\\
 &+&\left.
g_{\mathcal{H}W^+W^-}A_1^{\mathcal{H}}(\tau_W) 
+ \sum_s \beta_s Q_s^2 A_0^{\mathcal{H}}(\tau_s)
\right|^2,
\end{eqnarray}
where the loop functions are known and can be find, for instance, in~\cite{Djouadi:2005gj}.
The last term in eq.~(\ref{eq:diphoton}) represents
the contribution of the electrically charged scalar states, and we have
$\beta_s \equiv g_{\mathcal{H}H_sH_s^*}v/2m_s^2$. 

The electrically charged scalars affect the di-photon decay of both
 the new scalar resonance $H$ and the Higgs $h$ (the scalar couplings in eqs.~(\ref{eq:5},\ref{eq:3})
 for the Higgs boson can be found in~\cite{Hartling:2014zca}).
 The challenge is to explain 
 the di-photon signal strength observed by ATLAS and CMS 
 without introducing big deviation in the di-photon Higgs decay.

\subsection{Results: $\mu_H$ and $\Gamma_H$}

 %%%%%%%%%%%%%%%
\begin{figure}[!t!]
\centering
 \includegraphics[width = 0.45 \textwidth]{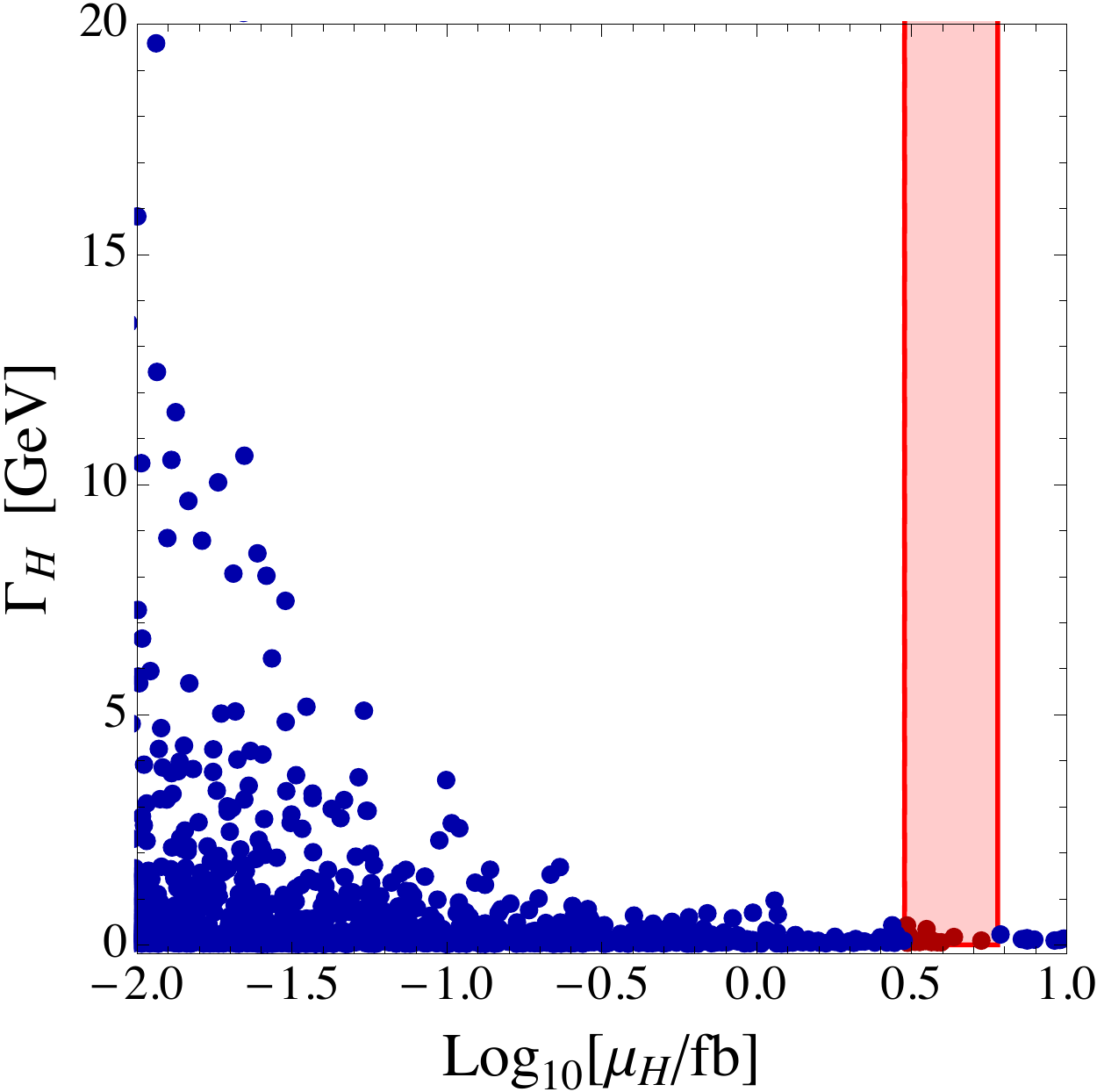}
\caption{\textit{
Result of the parameter scan in terms of total decay width $\Gamma_H$ versus
di-photon signal strength $\mu_H$ for the new scalar resonance at $m_H = 750$ GeV.
We mark in red the points where $\mu_H = [3-6]$ fb, as suggested by experimental data on the di-photon excess.
  }}
\label{fig:GammaRes}
\end{figure}
%%%%%%%%%%%%%%%%%%%%%%%%%%%%%%%%%%%%%

%Let us now discuss the results of our parameter scan.
%The most interesting findings are obtained considering negative mixing angles and scanning over $-M_2 \in (1,10^4)$ GeV.
In fig.~\ref{fig:GammaRes} we show the parameter scan in the plane $(\mu_H, \Gamma_H)$. There exists  a particular region of the scan where the model  reproduces a signal strength with size $\mu_H \sim O(1)$ fb. The red points in fig.~\ref{fig:GammaRes}, where $\mu_H$ is larger, correspond to the 
right-hand side of the allowed interval in $-M_2 \in (1,10^4)$ GeV.
In this range of values the scalar couplings in \eqs{eq:5}{eq:3}
are large, thus dominating the loop in $\Gamma_{\gamma\gamma}^H$.

%We will comment at the end of this section, in a more quantitative way, about the size of these couplings.

 %%%%%%%%%%%%%%%
\begin{figure}[!t!]
\centering
 \includegraphics[width = 0.45 \textwidth]{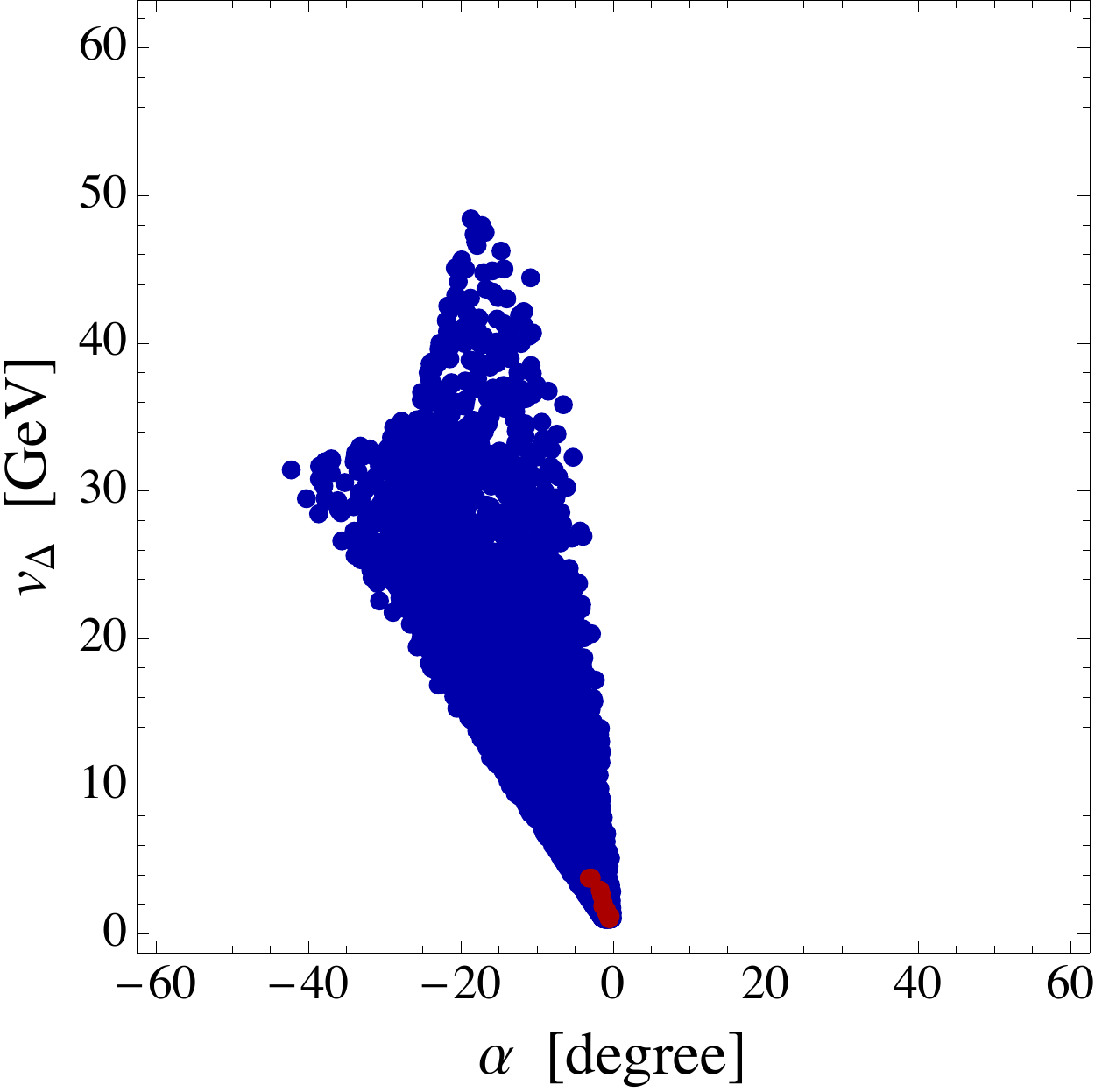}
\caption{\textit{
Result of the parameter scan in terms of mixing angle $\alpha$ versus
triplet VEV $v_{\Delta}$. 
We superimpose the analysed points to the region allowed by Higgs coupling measurements. The constraint from the EW parameter $S$ on  the scan is shown in green (1-, 2- and 3-$\sigma$ confidence level regions correspond to lighter shades).
  }}
\label{fig:Mixing}
\end{figure}
%%%%%%%%%%%%%%%%%%%%%%%%%%%%%%%%%%%%%

 %%%%%%%%%%%%%%%
\begin{figure}[!t!]
\centering
 \includegraphics[width = 0.45 \textwidth]{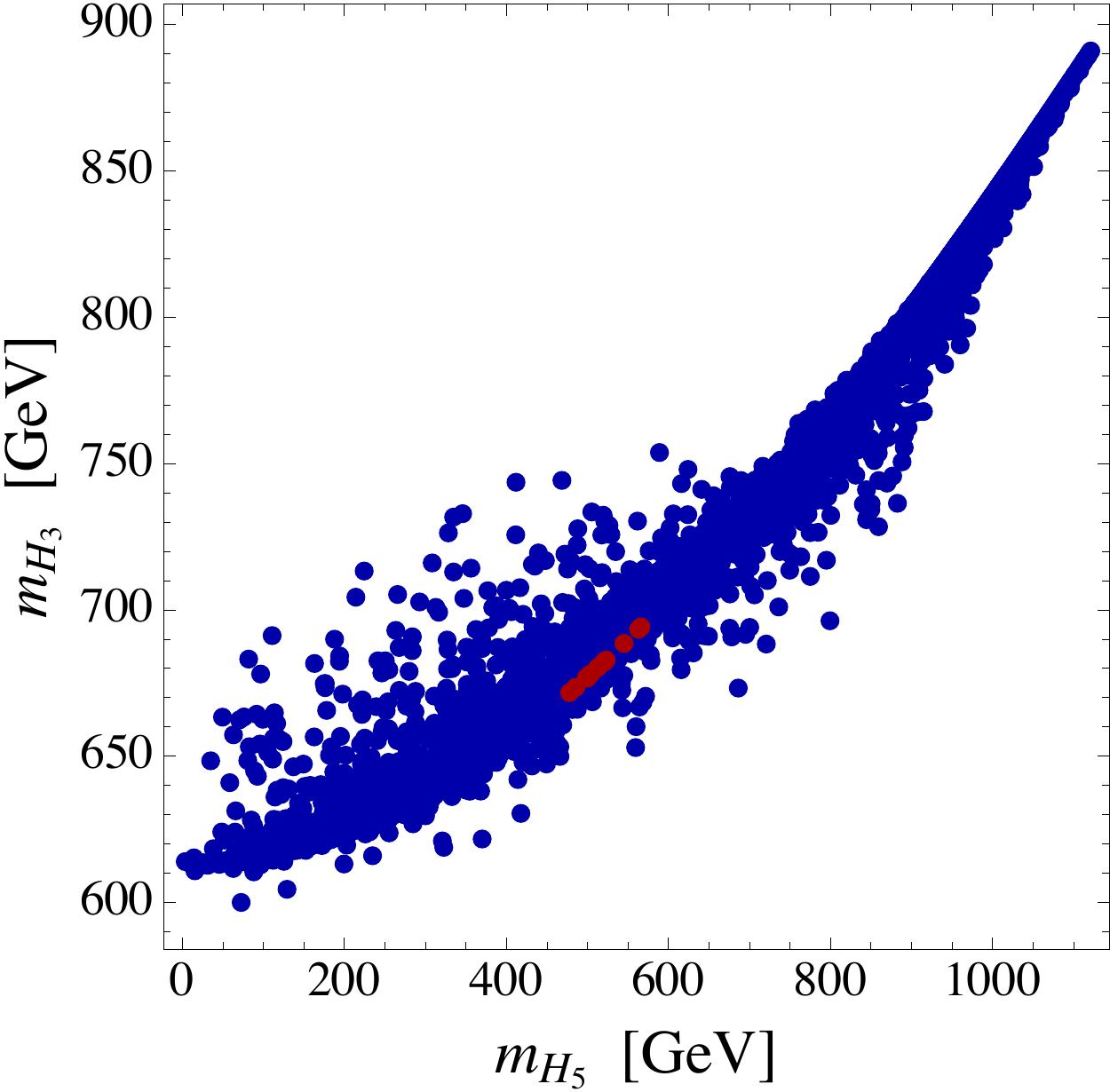}
\caption{\textit{
Result of the parameter scan in terms of the custodial triplet versus
quintuplet masses, $m_{H_{5,3}}$.
  }}
\label{fig:Spectrum}
\end{figure}
%%%%%%%%%%%%%%%%%%%%%%%%%%%%%%%%%%%%%

In fig.~\ref{fig:Mixing} we recast the parameter scan in the plane $(\alpha,v_{\Delta})$. The yellow contours agree with \cite{chiang}. 
We see that 
points where  $\mu_H \sim O(1)$ fb correspond to small and negative mixing angle, $\alpha \sim -3^{\circ}$
and triplet VEV $v_{\Delta} \lesssim 20$ GeV. 
In this region the dominant contribution to the production cross section is given by $\gamma\gamma$F.
Production  by means of  ggF and VBF contributes up to $20$\%.
As a consequence, the tension between the di-photon
excess observed at $\sqrt{s} = 13$ TeV and the absence of such signal in the dataset at $\sqrt{s} = 8$ TeV
is alleviated. The production cross section via ggF---going from $\sqrt{s} = 13$ TeV to $\sqrt{s} = 8$ TeV---is reduced by the factor 
$\sigma({\rm ggF}\to H)_{13\,{\rm TeV}}/\sigma({\rm ggF}\to H)_{8\,{\rm TeV}} = 4.693$ while
the production cross section via $\gamma\gamma$F is reduced by a factor of $2$.
These scaling factors make the di-photon excess at $\sqrt{s} = 13$ TeV consistent with the bound extracted from the $\sqrt{s} = 8$ TeV dataset.

In fig.~\ref{fig:Spectrum} we recast the parameter scan in the plane $(m_{H_5},m_{H_3})$.
Points where  $\mu_H \sim O(1)$ fb correspond to $m_{H_5} \sim 400-600$ GeV, $m_{H_3} \sim 650-700$ GeV.
This feature is expected because for these values the corresponding loop in the di-photon decay amplitude of $H$ is maximised.

%%%%%%%%%%%%%%%%%%%%%%%%%%%%%%%%%%%%%%
\begin{figure*}[!htb]
\minipage{0.33\textwidth}
  \includegraphics[width=.9\linewidth]{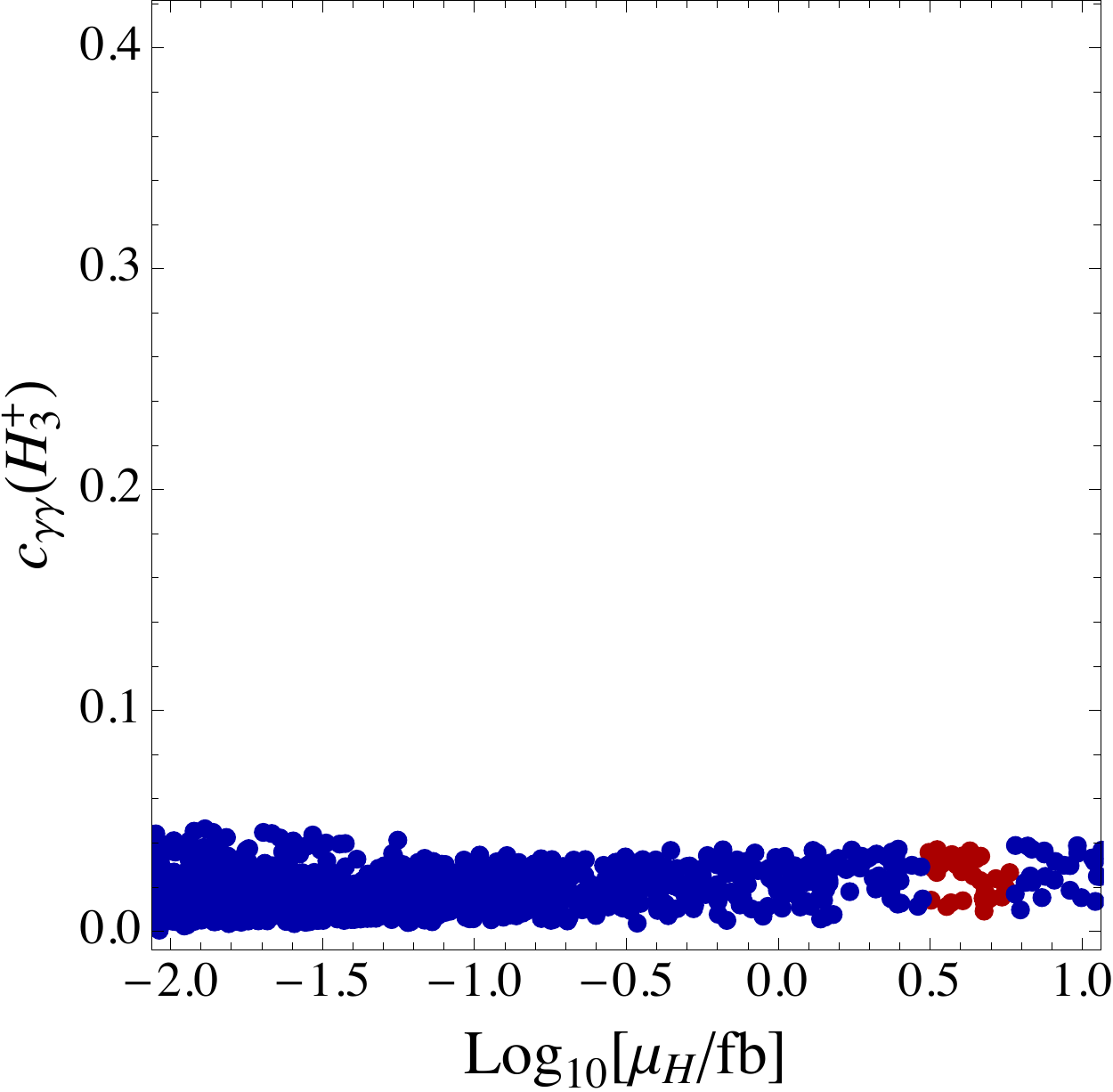}
\endminipage\hfill
\minipage{0.33\textwidth}
  \includegraphics[width=.9\linewidth]{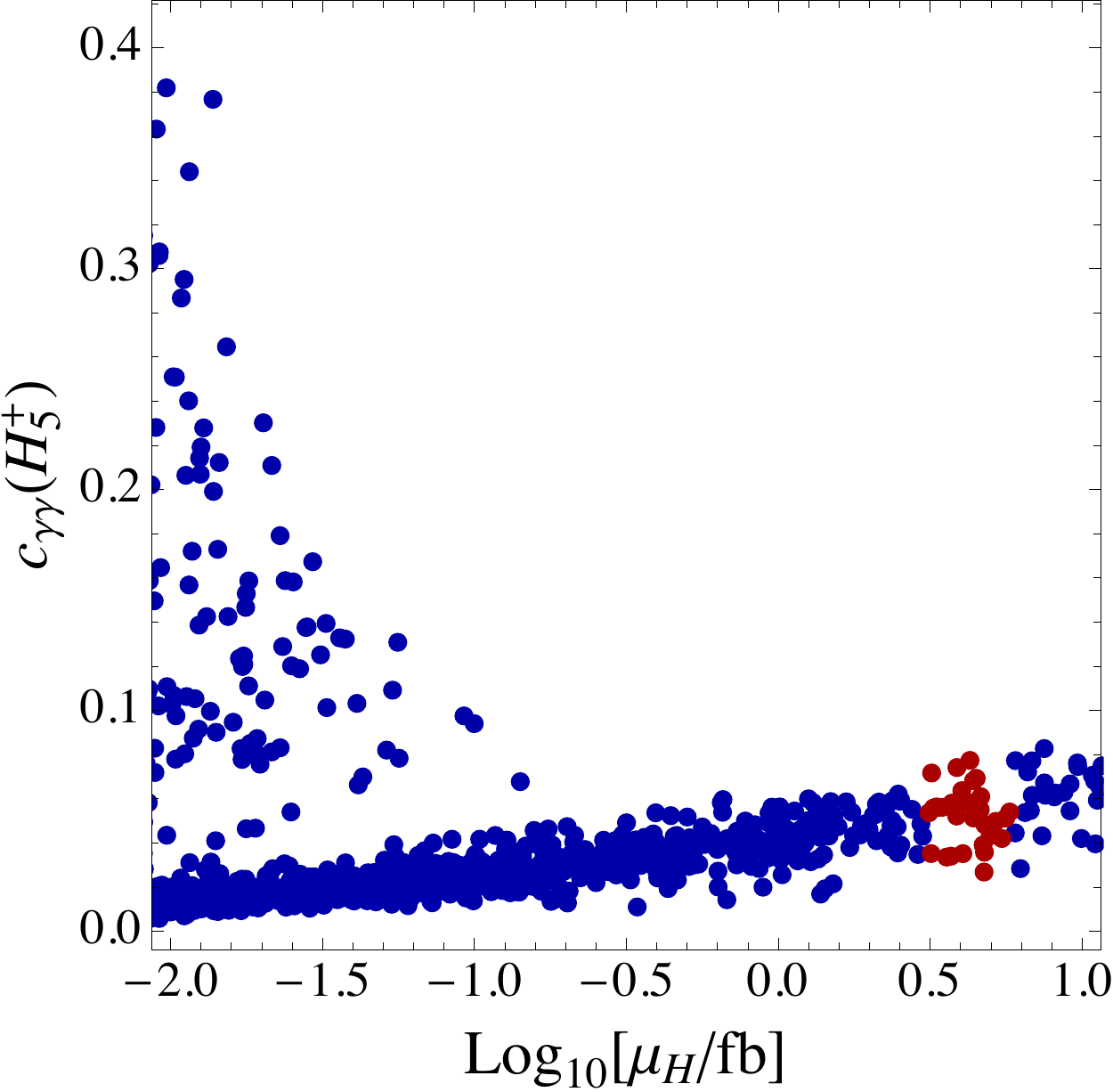}
\endminipage\hfill
\minipage{0.33\textwidth}
  \includegraphics[width=.9\linewidth]{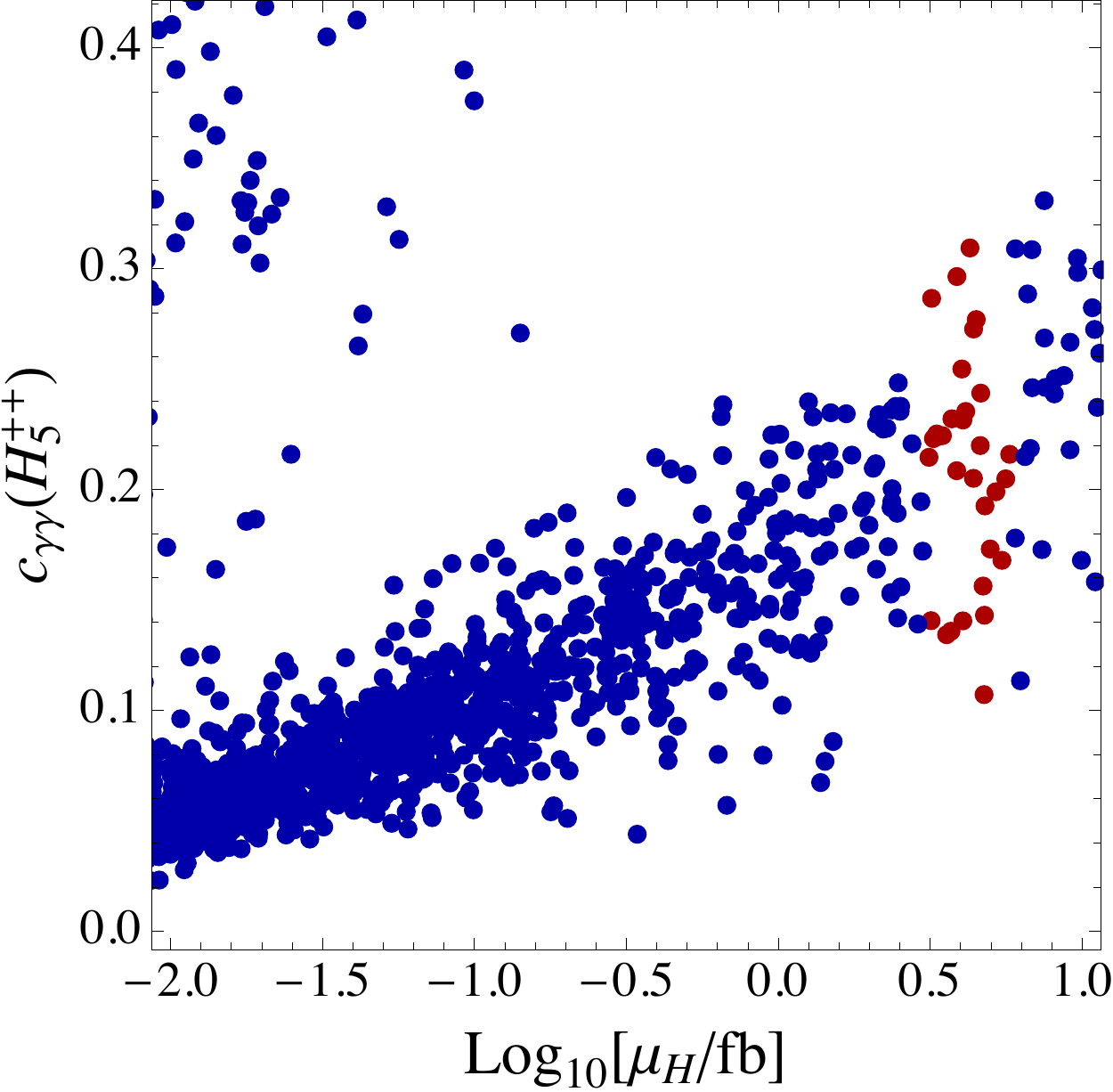}
\endminipage
 \caption{\textit{
Result of the parameter scan in terms of the Wilson coefficient in eq.~(\ref{eq:Wilson}).
  }}
 \label{fig:Couplings}
\end{figure*}
%%%%%%%%%%%%%%%%%%%%%%%%%%%%%%%

The explanation of the di-photon excess in the context of the GM model predicts the presence of additional 
light scalar degrees of freedom, including the doubly-charged state $H_{5}^{++}$.
Notice that tree-level decays of $H$ into triplet or quintuplet scalar states are not kinematically allowed at the red points of the scan.
 The characteristic phenomenology~\cite{Englert:2013wga,chiang} of these scalar states represents a signature of the model.

We checked that the model, for the chosen choice of parameter values,  is consistent with other searches for resonant production 
of a pair of SM particles which constrain the tree-level decay modes of $H$~\cite{Knapen:2015dap}. 

As it can be seen in fig.~\ref{fig:Mixing}---there is a moderate tension with the EW parameter $S$ for which the fit of the di-photon excess (the red dots) only agrees at the 3-$\sigma$ level. This is to be expected  given the presence of the additional charged new states.

The Higgs scaling factor $\kappa_{\gamma}$ is defined as the ratio between the loop-induced $h \to \gamma\gamma$ coupling 
in the GM model with respect to that of the SM. At the red points in fig.~\ref{fig:GammaRes},
we find $0.8 \lesssim \kappa_{\gamma} \lesssim 1.2$. The presence of such deviation 
is consistent with the present experimental bound~\cite{ATLAS_CMS}.

We find that the other two neutral scalars, $H_3^0$ and $H_5^0$ give a negligible contribution to the di-photon cross section. 

Concerning the total decay width $\Gamma_H$,
points where  $\mu_H \sim O(1)$ fb correspond to $\Gamma_H \sim 1$ GeV.
The value of the total decay width suggested by data represents at the moment 
the most controversial aspect of the di-photon excess.
Since the typical di-photon invariant mass resolution at $750$ GeV is estimated to be around $10$ GeV, 
it is natural to expect a large total decay width, $\Gamma_H \lesssim  40$ GeV.
At this stage of the experimental analysis no conclusive statements can be made, and the value $\Gamma_H \sim 1$ GeV
is perfectly consistent with the data. However, if large values of  $\Gamma_H$ are confirmed by future analysis, 
an explanation of the di-photon excess
in terms of weakly coupled theories will be disfavored.

\subsection{Perturbative reliability}

The result above is qualitatively different with respect to both the case in which the resonance is  not taking part in the EW symmetry breaking (and 
 one is forced to introduce additional electrically charged vector-like fermions to boost both production cross section and di-photon decay) and the 2HDM (in which 
the condition $\mu_H \sim O(1)$ fb requires unrealistically large Yukawa couplings).
In our scan, all the dimensionless couplings of the GM model are kept within the perturbative regime.

This point is better understood in terms of  the overall size of the di-photon decay induced by the loop of scalar particles.
In full generality, we can consider the effective Lagrangian 
\begin{equation}\label{eq:eff}
\mathcal{L}_{\rm eff} = \frac{e^2}{4v}c_{\gamma\gamma}HA_{\mu\nu}A^{\mu\nu}~,
\end{equation}
with $A_{\mu\nu}$ the usual photon field strength.
The effective operator in eq.~(\ref{eq:eff}) induces 
the di-photon decay
\begin{equation}
\Gamma_{\gamma\gamma}^{(H)} = \frac{c_{\gamma\gamma}^2 e^4 m_H^3}{64\pi v^2}~.
\end{equation}
We can recast, for illustrative purposes, the scalar loop contribution in eq.~(\ref{eq:diphoton})
in terms of the Wilson coefficient $c_{\gamma\gamma}$. 
Approximating for simplicity the scalar loop function as $A_0(\tau)\sim -1/3$, 
we find
\be\label{eq:Wilson}
c_{\gamma\gamma}(s) = \left[
\frac{\beta_s^2 Q_s^4}{36(4\pi)^2 \pi^2}
\right]^{1/2}\ \quad s=H_5^{+},\,H_5^{++},\,H_3^+\,.
\ee
In fig.~\ref{fig:Couplings} we show the typical size of these coefficients in our parameter scan.
The typical size is $c_{\gamma\gamma}(s)\sim 0.05$. The only exception 
is $c_{\gamma\gamma}(H_5^{++})$, which can reach values $c_{\gamma\gamma}(H_5^{++})\lesssim 0.4$ (due to the large electric charge, $Q_{H_5^{++}}^4 = 16$).

%%%%%%%%%%%%%%%%%%%%%%%%%%%%%%%%%%%%%%%%%%%%%%%%
%\vskip 1.3em
\acknowledgments
MF thanks  SISSA and the Physics Department at the University of Trieste  for the hospitality. 
We thank Michele Pinamonti, Florian Staub and Alberto Tonero for important discussions and advices.

%%%%%%%%%%%%%%%%%%%%%%%%%%%%%%%%%%%%%%

%%%%%%%%%%%%%%%%%%%%%%%%%%%%%%%%%%%%%%%%%%%%%%

\end{document}